# A high resolution multi-phase clock Time-Digital Convertor implemented on Kintex-7 FPGA

**X. Dong[a,b], C. Ma[c,1], X. Zhao[c], X. Li[c], Z. Huang[c]**

a. *Key Laboratory of Far-shore Wind Power Technology of Zhejiang Province, Hangzhou, 311199, China*
b. *School of Earth and Space Sciences, University of Science and Technology of China, Hefei, 230026, China*
c. *Minfound Medical System Co. Ltd. Dongshan Road, Shaoxing, 312000, China*

*E-mail*: macong@mail.ustc.edu.cn

ABSTRACT: Time-digital Converter (TDC) aims to measure the arrival time of the leading edge of the pulse signal. Our recent work presented a high resolution multi-phase TDC based on the Kintex-7 Field Programmable Gate Array (FPGA) device. A simple I/O tile based circular input buffer is employed to oscillate the input signal periodically, and then a multi-phase TDC based on ISERDES core with a 625 ps bin size is used to accomplish the multiple measurements for getting higher resolution performance. In this paper, the design concept, architecture, as well as kernel implementation considerations are all discussed. To evaluate the TDC's performance, we built a verification system based on Kintex-7 FPGA. Initial test results indicate that the TDC's effective bin size is successfully reduced from 625 ps to 78.125 ps, and the measured dual-channel time resolution is better than 35 ps RMS.

## Contents

## 1. Introduction

Nowadays, high-resolution time-to-digital converter (TDC) is widely applied in many measurement systems, such as particle detectors, medical imaging equipment, laser radar and aerospace systems [1-4]. Meanwhile, Field Programmable Gate Array (FPGA) is commonly used to accomplish digital algorithm or data transmission function in these systems. Therefore, it's much more cost-effective and integration-effective to employ a TDC based on FPGA (FPGA-TDC) rather than the external Application Specific Integrated Circuit (ASIC) chip [5]. As yet, the invented various FPGA-TDCs are mainly based on two strategies. One is to employ a series of flip-flops with the same clock to sample the tapped signal delayed by the internal delay chain resources (such as carry chain) [6]. This approach can achieve a low timing error (< 10 ps RMS) performance [7], but it is pretty hard to accomplish due to the dispersion and temperature drift of the delay chain. Generally, a complex calibration mechanism is necessary [8].

The other strategy is to employ several flip-flops with multi-phase clocks to sample the input signal [9]. The timing error for this approach is mainly determined by the frequency and phase shift of the sampling clocks, which can be synthesized by the Phase Locking Loop (PLL) in FPGA and are basically stable. This approach is also widely applied. For instance, Z. Yin achieved a multi-phase TDC with an around 363 ps RMS timing error in 2012 [10]. Mark D. Fries from General Electrical Medical Systems also produced a similar TDC with around 455 ps RMS timing error [11]. In 2016, Yonggang Wang successfully implemented a 256-channel high-resolution multi-phase clock TDCs with a resolution better than 56.2 ps RMS [12]. However, limited by the intrinsic properties of the device, it is pretty hard to achieve an ultra-resolution goal by just increasing the frequency or decreasing the phase shift of the sampling clocks.

In our recent work, an approach is presented to decrease the timing error of multi-phase TDC. A simple I/O tile based launcher is employed as a circular input buffer to oscillate the input signal periodically, and then a multi-phase TDC based on ISERDES core with a 625 ps bin size is used to achieve the multiple measurements for getting higher resolution performance. The improved TDC beyond the clock sampling bin size limitation doesn't need to implement ultrahigh frequency sampling clocks with lots of phases, which reduces the clocking resource usage of FPGA device. The paper is organized as follows: Section-2 introduces the structure of an normal ISERDES based TDC and its limitations; Section-3 presents a circular input buffer to increase the timing resolution, including the design concept, architecture and the kernel design considerations; the initial electronics test results of the TDC are shown in Section-4; finally, in section -5, we summary the paper and look forward to the future work.

## 2. Multi-phase TDC based on ISERDES and its limitations

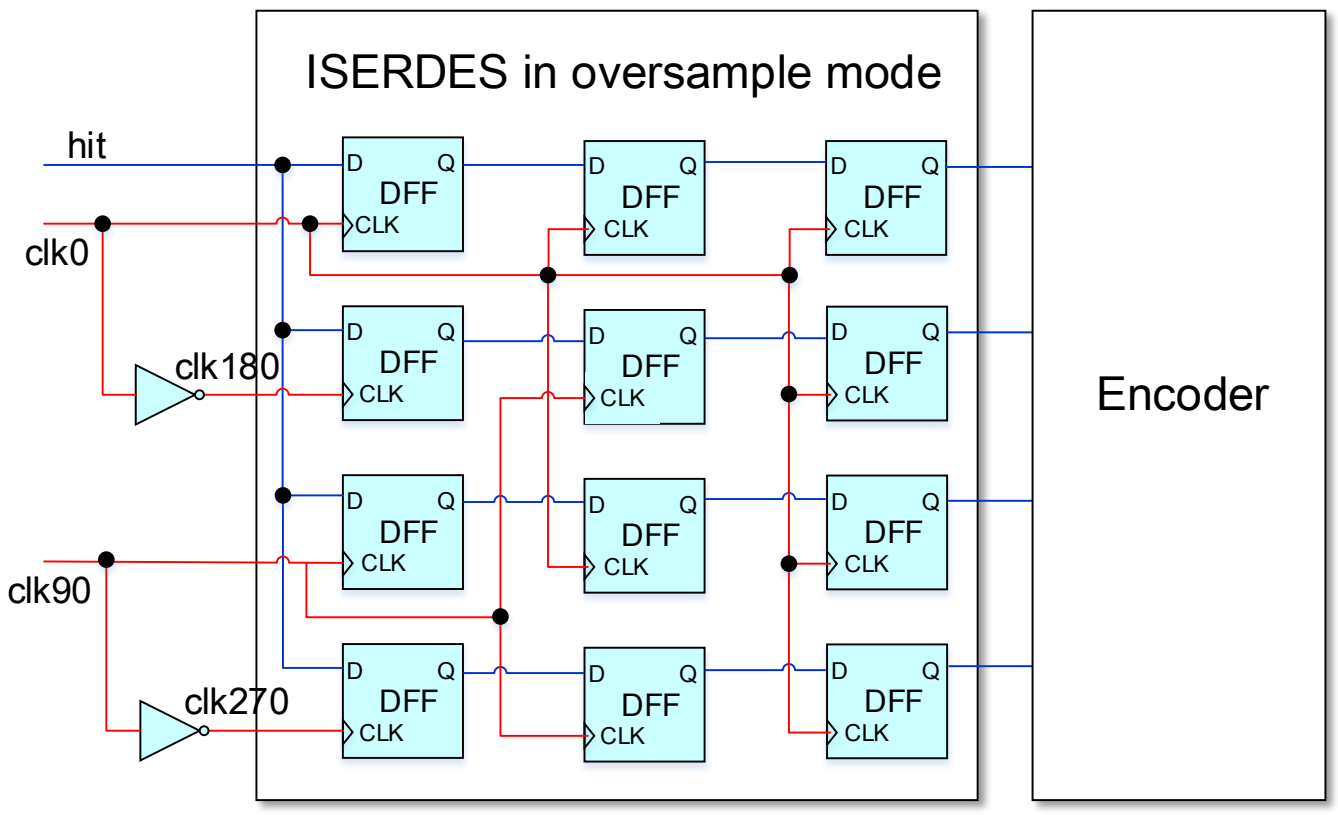

**Figure 1** Multi-phase TDC based on ISERDES core

In 2014, Tian. X presented a FPGA I/O tile based TDC using ISERDES core (named iTDC in this paper), as shown in Figure 1 [13]. The ISERDES core configured in oversample mode has tree-stage flip-flops structure and can be implemented as a multi-phase TDC. PLL in FPGA device generates two high frequency clocks with 90° phase shift (clk0, clk90), which are inverted by the non-gates to generate other two 90° phase shift clocks (clk180, clk270). These multi-phase clocks are respectively connected to the CLK, CLKB, OCLK and OCLKB ports of the ISERDES core to sample the input signal (hit). Therefore, the bin size of this iTDC ($bin$) is determined by the following equation:

$$bin = \frac{1}{f \times P} \qquad (1)$$

Where $f$ and $P$ is the frequency and the number of phases of the sampling clocks.

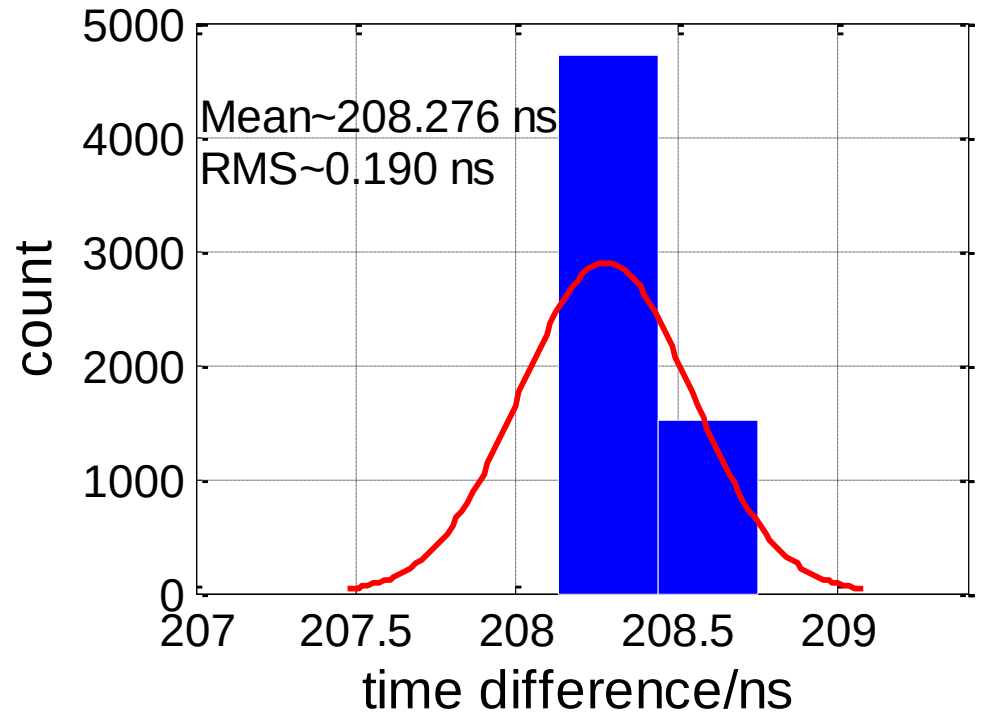

**Figure 2** Typical dual-channel time resolution test results of iTDC

Compared with the traditional multi-phase clock TDC composed with the discrete flip-flops, this architecture reduces the performance deterioration caused by the difference between the routing delays and the difficulty of the placement constraints. Nevertheless, it is still very difficult to reduce the bin size. A typical dual-channel time resolution test result is shown in Figure 2, which demonstrates the limitation in time resolution. Even for 2 parallel ISERDES core with 800 MHz sampling clocks, the bin size can be only reduced to around 156 ps. The resolution of this level of TDC is not sufficient in some high precision measurement systems, such as Time-to-Flight (TOF) Positron Emission Computed Tomography (PET) equipment [13]. In one word, the resolution performance limits its scope of application and fields. Our recent work focuses on trying a new scheme to improve its time resolution performance, which is presented in the next section.

## 3. Improved TDC using circular buffer

### 3.1 Improved TDC architecture

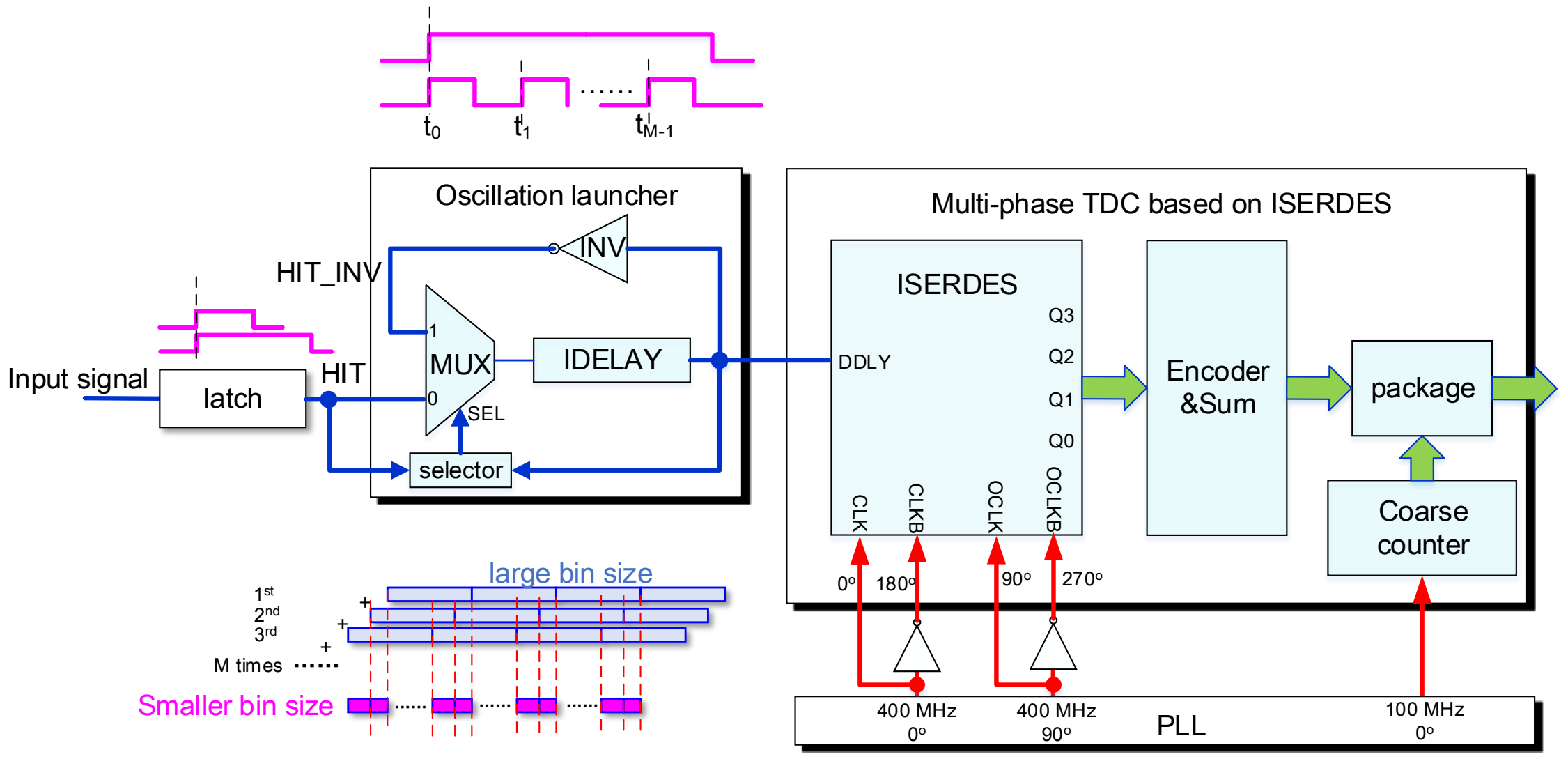

**Figure 3** Architecture of TDC

The architecture of the improved TDC is shown in Figure 3. Internal PLL generates one 100 MHz system clock (period: $T_{CLK}$) as the sampling clock of a 40-bit coarse timer and four 400 MHz ISERDES core sampling clocks to achieve time interpolation measurement. Therefore, the bin size of iTDC is 625 ps and the time measurement dynamic range is larger than 3 hours.

In order to reduce the bin size and increase the time resolution performance, a simple I/O tile based oscillation launcher is employed and next describes its working steps: The leading edge of the coming input signal (positive impulse, *HIT*) is latched by the flip-flops, and then fed into the $0^{th}$ input data port of a multiplexer ($MUX$, propagation delay: $t_{MUX}$), whose output is delayed by the following IDELAY core (propagation delay: $t_{IDELAY}$, user defined). The frequency of the IDELAY reference clock ($F_{ref}$) is 300 MHz, so the adjustable tap is around 52 ps ($\frac{1}{64F_{ref}}$). Since the $MUX$ is power on reset to *HIT* ($SEL$ is '0'), after around $t_{MUX}+t_{IDELAY}$, the *HIT* is fed into the DDLY input ports of the ISERDES core to accomplish the first leading edge timing measurement. Meanwhile, the delayed *HIT* is also inverted by a non-gate (*INV*) and then fed into the $1^{st}$ data port of the $MUX$. Therefore, the $MUX$ is switched to '1' after the delayed *HIT* pulls the $SEL$ up. Similarly, after $t_{MUX}+t_{IDELAY}$, the $MUX$ is switched to '0' again and this forms an simple oscillation launcher (oscillating period: $T_{OSC}$). The number of cycles per *HIT* (*M*) are user assignable. Internal logic (selector) counts the cycles of the oscillator when a hit arrives and switches $MUX$ back to *HIT* by $SEL$ at the end of the $M^{th}$ cycle of the oscillating. Then the TDC is ready for the new incoming signal.

The output thermometer code of the iTDC can be further encoded to binary code by the following pipelined encoder, and then sum the measurement results of each time ($t_i$, $0<i<M$) up as the final fine time information ($t_{fine}$). Moreover, $T_{OSC}$ is designed larger than 5 ns so that in one system clock period ($T_{CLK}$, 10 ns), no more than two leading edges of the $T_{OSC}$ can be recorded. The $t_{fine}$ calculation equation is shown in Equation-2, where $m(i)$ is the $T_{CLK}$ compensation value of the $i^{th}$ measurement. By multiple measurements, the improved TDC will get smaller bin size and higher resolution performance than the plain iTDC.

$$t_{fine} = \sum_{i=0}^{i=M-1}(t_i + m(i) \times T_{CLK}) \tag{2}$$

where $m(i)$ is 0 when two edges are detected or 1 when only one edge is detected in one cycle.

In addition, $T_{OSC}$ can be calculated by the following equation:

$$t_{i+1} - t_i = T_{OSC} - m(i) \times T_{CLK} \tag{3}$$

## 3.2 Implementation considerations

For the improved TDC, it is important to control the $T_{OSC}$ which affects the resolution performance. $T_{OSC}$ mostly consists of the propagation delay of the *INV* component ($t_{INV}$), $t_{MUX}$, $t_{IDELAY}$ and the routing propagation delay ($t_{routing}$) from *MUX* to IDELAY ($t_{MUX\text{-}IDELAY}$), from IDELAY to *INV* ($t_{IDELAY\text{-}INV}$), as well as from *INV* to *MUX* ($t_{INV\text{-}MUX}$). Placement constraints should be carefully applied to locate *INV* and *MUX* to a specific position within the time interpolating chain to avoid unpredictable routing delay. An approximation of the oscillating period is given as:

$$T_{OSC} \sim 2 \times (t_{IDELAY} + t_{INV} + t_{MUX} + t_{routing}) \tag{4}$$

where $t_{routing} = t_{MUX-IDELAY} + t_{IDELAY-INV} + t_{INV-MUX}$.

Since multiple measurement for the delayed input signals is equivalent to employ multiple iTDCs with evenly phase-shift sampling clocks, the best case is that the several leading edges of the oscillating *HIT* just fall in the bin size of the iTDC with an equal interval. In other words, the remainder of $T_{OSC}$ divided by *bin* of iTDC happens to be *bin*/*M*:

$$M = \frac{1}{T_{OSC} MOD \frac{1}{f \times P}} \times \frac{1}{f \times P} \quad (5)$$

Although it is a bit difficult to meet the above harsh condition, we can have the aid of post-implementation timing simulation by the Vivado platform and actual tests of the $T_{osc}$ by the oscilloscope to optimize the placement constraints and the IDELAY delay value.

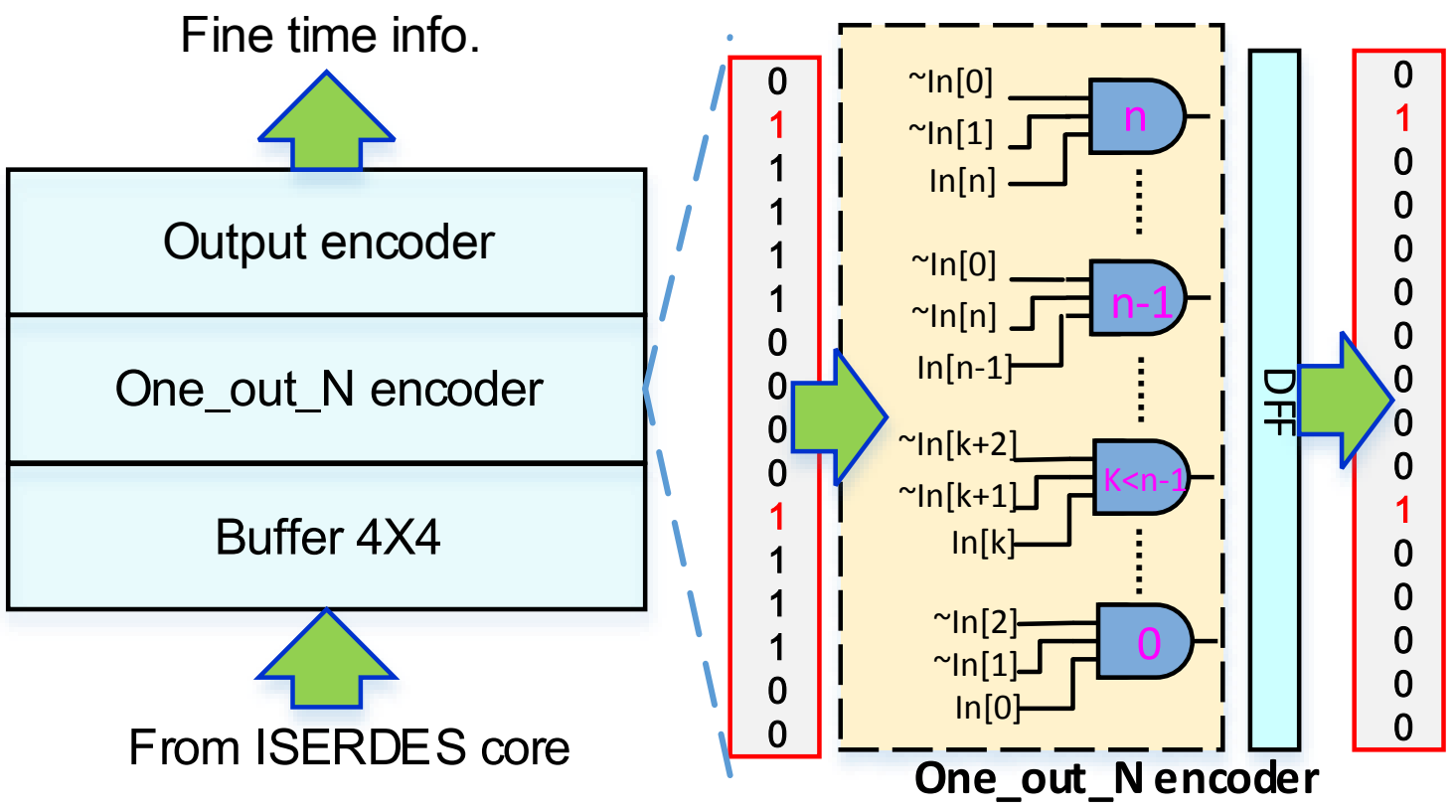

**Figure 4** structure of the Encoder

It must also be assured is that the pipelined encoder can handle the high-speed data stream from the ISERDES and encode them to the binary codes efficiently and reliably. The structure of the encoder is shown in Figure 4. First the outputs of the ISERDES (@ 400 MHz) are buffered by flip-flops array (4×4) , and then fed into an one_out_N encoder [14]. The one_out_N encoder consists of 16 3-ports NAND gates to locate the '0-1' transient position which represents the leading edge arrival time of *HIT*, and the following output encoder can easily convert it to a binary code. Reference [14] shows more details about this. Finally, the coarse and fine time information are packed and buffered.

## 4. Performance

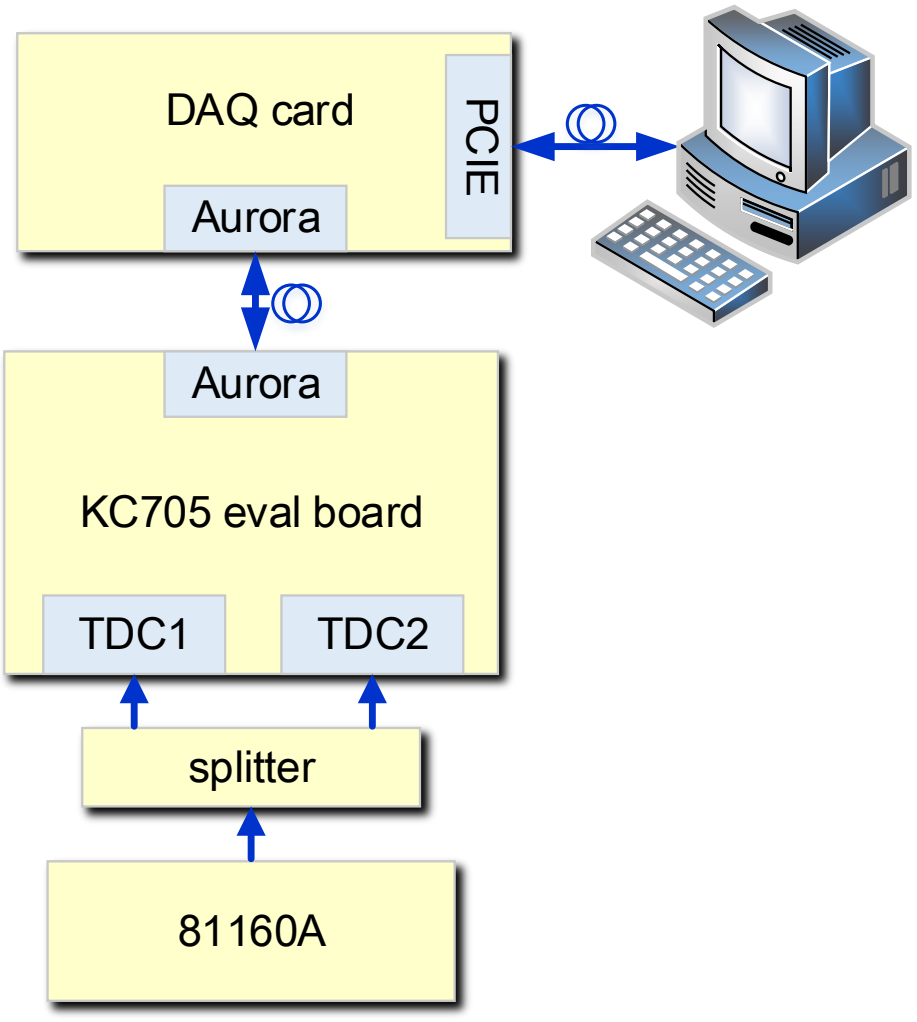

**Figure 5** TDC evaluation platform

To evaluate the TDC's performances, we built a verification system based on Kintex-7 FPGA, as shown in Figure 5. The output of the pulse generator (81160A, from Keysight Inc.) is split and fed into two TDC channels, which are implemented in a Xilinx evaluation board (KC705) based on Kintex-7 FPGA (XC7K325T-2FFG900). The measurement results are transmitted to a data acquisition (DAQ) card via an optical fiber, and then sent to PC via the peripheral component interconnect express (PCIE) interface.

### 4.1 $T_{osc}$ period histogram

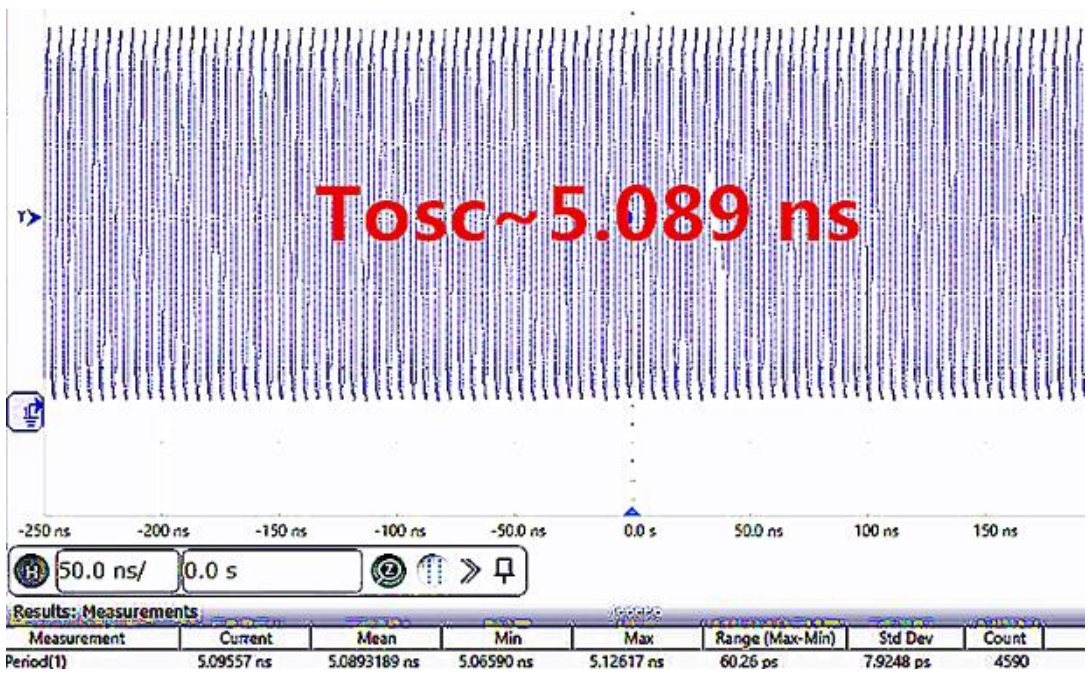

**(a)**

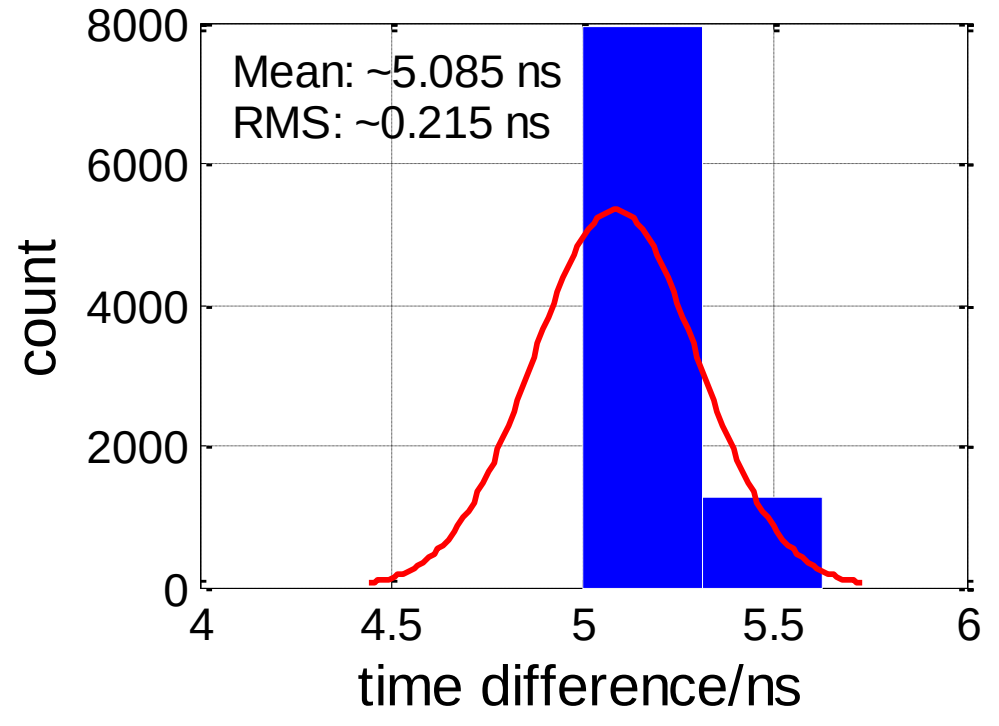

**(b)**

**Figure 6** $T_{OSC}$ period test results. (**a**: by oscilloscope **b**: by iTDC)

In order to verify the correctness and stability of the encoder, according to the Equation-3, the $T_{OSC}$ can be measured by the iTDC. The test result, as shown in Figure 6(b), indicates that the $T_{OSC}$ is around 5.085 ns with a 215 ps RMS jitter. The mean is very close to the oscilloscope test result, which is around 5.089 ns with a 7.9 ps RMS jitter.

## 4.2 Effective bin size and nonlinearity

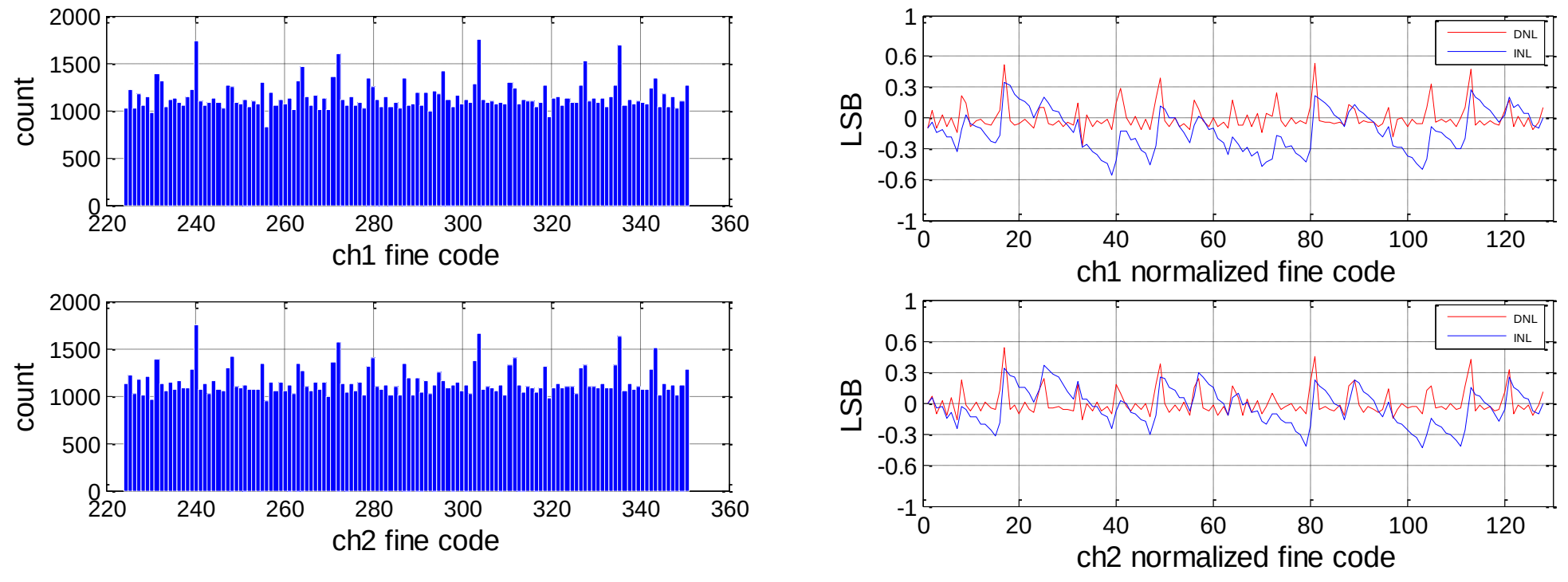

**(a)** **(b)**

**Figure 7** The fine code distribution map and the nonlinearity test results of two TDC channels (**a**: The fine code distribution map. Fine code is in the range of 224~351. **b**: The nonlinearity test results. The fine code is normalized to 1~128)

According to the code density method [15], there is an equal probability of the hit signal in any position of the clock period, and the relative fractions of the hit signal detected in a TDC bin reflect the TDC bin width. Figure 7 (a) is the typical measured fine code distribution map of two channels. The fine code is from 224 to 351, and the effective bin size is 78.125 ps, which meets the design expectations quite well. Moreover, as shown in Figure 7 (b), the differential nonlinearity (DNL) and integral nonlinearity (INL) lie in range between -0.28 LSB ~ +0.53 LSB and -0.56 LSB ~ +0.38 LSB without any nonlinearity correction, respectively.

## 4.3 Time resolution performance

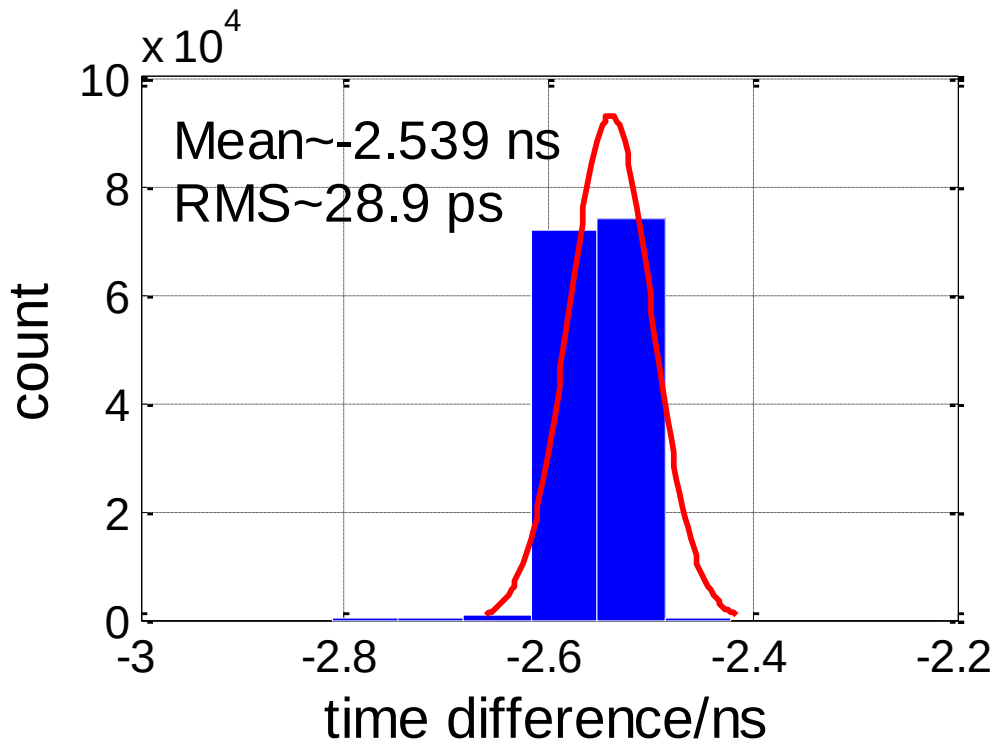

**Figure 8** A typical dual-channel time resolution test results of the improved TDC

We use the cable delay method to assess the time resolution performance. The output signal is split and imported to two TDC channels. Considering the situation where the time results of these two channels are not interrelated, the single-channel time resolution can be obtained by dividing by the RMS value of the time difference by $2^{1/2}$ [16]. A typical time difference histogram between two channels is shown in Figure 8. Test result indicates that the dual-channel time difference is around -2.539 ns MEAN (-32.5 LSB) and the timing error is around 28.9 ps RMS (0.37 LSB). Obviously, this resolution performance is much better than the plain iTDC.

By adjusting the input interval delay of the two channels, two longer time interval sweep tests are also conducted. One of the tests has a sweep range over 2 ns with a 100 ps step, and another is over 2 us with a 50 ns step. Both the time mean and jitter of the interval are calculated. Test results indicate that the TDC works quite well over a large dynamic range and the time resolution is better than 35 ps RMS, as shown in Figure 9.

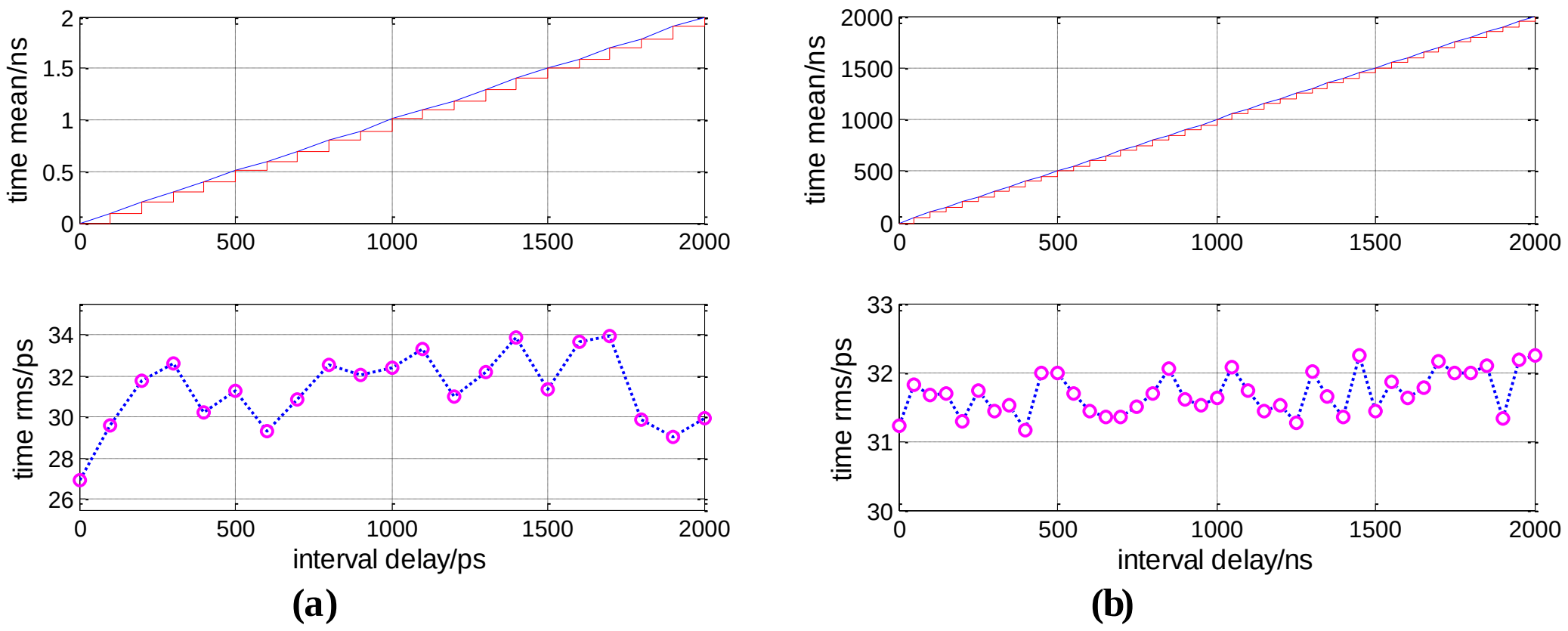

**Figure 9 time interval sweep test results**
**(a: test over a dynamic range of 2 ns with 100 ps step**
**b: test over a dynamic range of 2 us with 50 ns step)**

## 4.4 Dead time

As mentioned above, $T_{OSC}$ is controlled around 5.09 ns and $M$ is 8. Both the Encoder and buffer are all pipelined. So the dead time is less than 50 ns, which means the peak measurement throughput of the TDC can reach 20 M events per second.

## 4.5 Parameter comparison

Parameters of several types of TDC schemes are compared in Table 1. The improved multi-phase clock TDC has some advantages of resolution and resouce usage.

Table 1 Parameters of Several typical TDC Architectures

| | One carry chain TDC[17] | One ISERDES based multi-phase clock TDC[13] | The improved multi-phase TDC |
|---|---|---|---|
| Bin width | ~30 ps | ~156 ps | 78.125 ps |
| Typical RMS | ~10 ps | ~56 ps | ~35 ps |
| LUT useage | 1621 | 109 | 199 |
| Flip-flop usage | 1621 | 238 | 347 |

# 5. Discussion and Conclusion

## 5.1 Discussion

As mentioned above, a high-resolution multi-phase clock TDC prototype is presented. Nevertheless, further in-depth work should be done to evaluate and perfect it.

1). Multi-channel tests at different ambient temperature. As discussed, the $T_{OSC}$ should be carefully controlled and it has difference among channels, which leads to the timing error dispersion. Besides, the I/O tile temperature drift should also be evaluated. Therefore, more channels should be tested at different ambient temperatures, which will be reported in future.

2). Further improve the resolution. In this paper, $M$ is optimized to 8 to achive a 7-bit TDC. If the $T_{OSC}$ is futher optimized, it is possible to set a larger $M$ and get a smaller bin size. Besides, in our current design, the adjustable step of the IDELAY is about 50 ps. Maybe a very short carry chain resource (smaller adjustable step) can be employed to control the $T_{osc.}$

## 5.2 Conclusion

In summary, we proposed an improved ISERDES based TDC architecture. A simple I/O tile based launcher is introduced to oscillate the input signal periodically, and then a multi-phase TDC based on ISERDES core is used to achieve the multiple measurement. In contrast to the plain ISERDES based TDC, this architecture has better time resolution performance, beyond the sampling clocks bin size limitation. Initial test results indicate that the TDC's effective bin size is successfully reduced from 625 ps to 78.125 ps, and without any nonlinearity correction the DNL and INL lie in range between -0.28 LSB ~ +0.53 LSB and -0.56 LSB ~ +0.38 LSB, respectively. The measured dual-channel time resolution is better than 35 ps RMS. The results show that multi-phase clock sampling method based TDCs may have a broader range of applications.

## References

[1] L.F. Kang, L. Zhao et al, *A 128-channel high precision time measurement module*, *Metrol. Meas. Syst* .**20** (2013) 275.
[2] Long Y , Guo-Dong L , Ru-Mei Z . *Design and implementation of front-end electronic system for TOF-PET. Electronic Design Engineering,* **12**(2019).
[3] Nissinen J , Nissinen I , Kostamovaara J . *An integrated CMOS receiver-TDC chip for mm-accurate pulsed time-of-flight laser radar measurements. IEEE International Instrumentation & Measurement Technology Conference.* Graz, Austria, 2012.
[4] Qin X , Feng C , Zhao L , et al. *Development of high resolution TDC implemented in radiation tolerant FPGAs for aerospace application, IEEE Real Time Conference (RT)*, California, 2012.
[5] E. Bayer and M. Traxler, *A high-resolution (< 10 ps RMS) 32-Channel Time-to-Digital Converter (TDC) implemented in a Field Programmable Gate Array (FPGA), 2010 IEEE-NPSS Real Time Conference,* Lisbon, 2010, pp. 1-5.
[6] Jian Song, Qi An and Shubin Liu, *A high-resolution time-to-digital converter implemented in field-programmable-gate-arrays. IEEE Transactions on Nuclear Science*, vol. 53, no. 1, pp. 236-241, Feb. 2006
[7] Q. Shen et al., *A 1.7 ps Equivalent Bin Size and 4.2 ps RMS FPGA TDC Based on Multichain Measurements Averaging Method. IEEE Transactions on Nuclear Science,* vol. 62, no. 3, pp. 947-954, June 2015.
[8] C. Hervé, J. Cerrai, T. Le Caër, *High resolution time-to-digital converter (TDC) implemented in field programmable gate array (FPGA) with compensated process voltage and temperature (PVT) variations. Nuclear Instruments and Methods in Physics Research Section A: Accelerators, Spectrometers, Detectors and Associated Equipment*, vol. 682, pp. 16-25, 2012
[9] Qi Z, Meng X , Li D , et al. *A high precision TDC based on a multi-phase clock. physics*, 2015.
[10] Yin Z J C, Liu S B, Hao X J, et aI. *A High-Resolution Timeto-Digital Converter Based on Multi-Phase Clock Implement in Field Programmable-Gate-Array, IEEE Real Time Conference (RT),* California, 2012
[11] Fries, Mark D., and John J. Williams. *High-precision TDC in an FPGA using a 192 MHz quadrature clock. IEEE Nuclear Science Symposium Conference Record,* Vol. 1, 2002.
[12] Y. Wang, P. Kuang and C. Liu, *A 256-channel multi-phase clock sampling-based time-to-digital converter implemented in a Kintex-7 FPGA, 2016 IEEE International Instrumentation and Measurement Technology Conference Proceedings,* pp. 1-5, Taipei, 2016.
[13] Tian Xiang, Lei Zhao, Xi Jin, etal. *A Multi-phase Clock Time-to-Digital Convertor Based on ISERDES Architecture, IEEE International Symposium on Field-programmable Custom Computing Machines,* 2014.
[14] Qi Shen, Lei Zhao etal. *A fast improved fat tree encoder for wave union TDC in an FPGA, Chinese physics C,* 2013(10).
[15] Yun-Fei J , Zhi-Peng Z , Meng-Qiang X U , et al. *Research on Non-Linear Calibration for Time to Digital Converter Based on Code Density Distribution, measurement & control technology,* 2015(1).
[16] J. Wang, S. Liu, L. Zhao, X. Hu and Q. An, *The 10-ps Multitime Measurements Averaging TDC Implemented in an FPGA, IEEE Transactions on Nuclear Science,* vol. 58, no. 4, pp. 2011-2018, Aug. 2011
[17] Wu, Jinyuan, and Zonghan Shi. *The 10-ps wave union TDC: ImprovingFPGA TDC resolution beyond its cell delay. IEEE Nuclear Science Symposium Conference Record*, 2008.